\documentclass[twocolumn, trackchanges]{aastex63}

\usepackage{graphicx}
\usepackage{apjfonts}
\usepackage{amsmath}
\usepackage[raggedright]{subfigure}
\usepackage{color}
\usepackage{hyperref}
\usepackage{acronym}
\usepackage[english]{babel}
\usepackage {lmodern}

\shorttitle{the influence of MB laws on LMXB evolution} 
\shortauthors{\sc Yang \& Li}

\begin{document}

\title{The influence of the magnetic braking laws on the evolution of persistent and transient low-mass X-ray binaries}

\author[0000-0001-5532-4465]{Hao-Ran Yang}
\author[0000-0002-0584-8145]{Xiang-Dong Li}

\affiliation{School of Astronomy and Space Science, Nanjing University, Nanjing 210023, China; lixd@nju.edu.cn}
\affiliation{Key Laboratory of Modern Astronomy and Astrophysics (Nanjing University), Ministry of Education, Nanjing 210023, China}


\begin{abstract} 

\noindent Swift J1858.6$-$0814 (hereafter J1858) is a transient neutron star low-mass X-ray binary (NS LMXB). There is controversy regarding its donor mass derived from observations and theoretical calculations. In this paper, we adopt seven magnetic braking (MB) prescriptions suggested in the literature and different metallicity $Z$ to simulate the evolution of the LMXB. Our results show that, employing the MB model proposed by \citet{2012ApJ...746...43R} ("rm12"), the Convection And Rotation Boosted ("carb") model \citep{2019ApJ...886L..31V}, as well as the Intermediate ("inter") and Convection-boosted ("cboost") models in \citet{2019MNRAS.483.5595V} can match (part of) the observational parameters of J1858 well. We then apply our method to other observed LMXBs and find that the "rm12" and "inter" MB laws are most promising in explaining transient LMXBs. In comparison, the simulations with the "cboost" and "carb" MB laws are more inclined to reproduce persistent LMXBs and ultra-compact X-ray binaries (UCXBs), respectively. Our results, though subject to computational and/or observational bias, show that it is challenging to find a unified MB law that applies to the NS LMXB sub-populations simultaneously, indicating our lack of understanding of the true MB law. In addition, we explore the influence of various MB laws on the magnitude of the bifurcation periods in LMXBs.
\end{abstract}


\section{Introduction}\label{s:intro}
Low-mass X-ray binaries (LMXBs) are systems comprised of either a black hole (BH) or a neutron star (NS) accreting material from a low-mass donor star. So far, there are about 350 known sources in the Galaxy, including 75 confirmed BH LMXBs or candidates, with the remainder being (potential) NS LMXBs \citep{2023A&A...675A.199A}\footnote{\href{http://astro.uni-tuebingen.de/~xrbcat/LMXBcat.html}{http://astro.uni-tuebingen.de/~xrbcat/LMXBcat.html}}. J1858 is among the confirmed NS LMXBs, firstly detected as a transient sources in October 2018 \citep{2018ATel12151....1K}. It has an estimated distance of $12.8^{+0.8}_{-0.6}$ kpc \citep{2020MNRAS.499..793B} and an orbital period of $21.3$ hrs with an eclipse duration of 4098 s \citep{2021MNRAS.503.5600B}. \citet{2022MNRAS.514.1908K} modeled the energy-dependent
eclipse profiles of J1858 in X-ray
energy bands and determined that the binary has an orbital inclination of $\sim 81^{\circ}$ and a mass ratio of $q\sim 0.14$, indicating a low-mass donor in the mass range $0.183\ \rm{M}_{\odot}\leq$ $M_{\rm d}\leq 0.372\ \rm{M}_{\odot}$. However,  the anomalous N V, C IV, Si IV and He II lines observed by \citet{2024MNRAS.527.2508C} indicate that the donor star of J1858 has undergone CNO processing, implying that the donor's initial mass is larger than $2\ \rm{M}_{\odot}$. By modeling the binary evolution the authors predicted that the present donor mass should fall within the range $0.5\ \rm{M}_{\odot}\leq$ $M_{\rm d}\leq 1.3\ \rm{M}_{\odot}$, which is in tension with the results suggested by eclipse modeling \citep{2022MNRAS.514.1908K}.

The evolution of LMXBs is influenced by various factors, encompassing initial parameters such as the donor mass, orbital period and metallicity. Additionally, the donor's response to mass and angular momentum loss, including gravitational radiation and magnetic braking (MB), also plays a vital role. For example, the value of the bifurcation period, which separates the converging from diverging binary systems \citep{1988A&A...191...57P,1989A&A...208...52P}, strongly depends on the strength of MB \citep{2009ApJ...691.1611M,2021ApJ...909..174D}. However, the precise physics underlying MB remains a subject of debate.

The predominant MB model was firstly introduced by \citet{1972ApJ...171..565S} and subsequently formulated by \citet{1981A&A...100L...7V} and \citet{1983ApJ...275..713R} for LMXBs, which has been extensively employed in stellar and binary evolution calculations. However, it was noted that the numerically calculated mass transfer (MT) rates using this model are approximately one order of magnitude lower than the observed values of some NS LMXBs \citep{2002ApJ...565.1107P}. In recent years, quite a few alternative MB models have been proposed \citep[e.g.,][]{2012ApJ...746...43R,2012ApJ...754L..26M,2019ApJ...886L..31V,2019MNRAS.483.5595V}, each of which can account for certain observations to varying degrees. However, a unified conclusion has yet to be achieved. For J1858, \citet{2024MNRAS.527.2508C} employed the Convection And Rotation Boosted recipe \citep{2019ApJ...886L..31V} and considered a solar abundance zero-age main sequence star for the evolution of the donor star. In this work, we attempt to simulate the LMXB evolution using various proposed MB models and metallicities to search if there are other (more) adequate models that can explain the observations. More broadly, the simulated results can be compared with observational properties (including the donor masses, orbital periods, and MT rates) of the current NS LMXB population. Furthermore,  exploring the influence of the MB recipes and metallicity on the bifurcation periods of LMXBs would also be of considerable interest.

The rest of the paper is organized as follows. In Section~\ref{s:method}, we introduce the binary evolution models used in this paper, including the MB models and other detailed prescriptions. The simulated results are presented in Section~\ref{s:result}, which are also compared with observations. Our discussion and conclusions are  given in Section~\ref{s:conclusion}.

\section{Binary evolution model}\label{s:method}
\begin{figure*}[t]
	\centering \includegraphics[width=\textwidth]{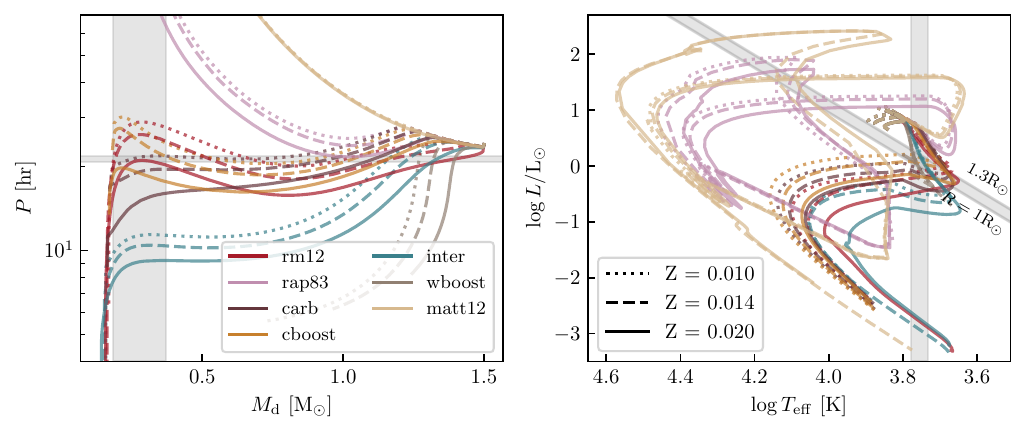}
	\caption{Example evolution for different metallicities $Z$ (denoted by different line styles) and MB laws (denoted by different colors) with initial donor mass $M_{\rm d0}=1.5\ \rm M_{\odot}$ and orbital period $P_0=1.5\ \rm d$. The left and right panels represent the evolution in the $M_{\rm d}-P$ plane and the HR diagram, respectively. The shaded areas are the observed (or inferred) values of J1858.}
	\label{fig:example}
\end{figure*}
\begin{figure*}[t]
	\centering \includegraphics[width=\textwidth]{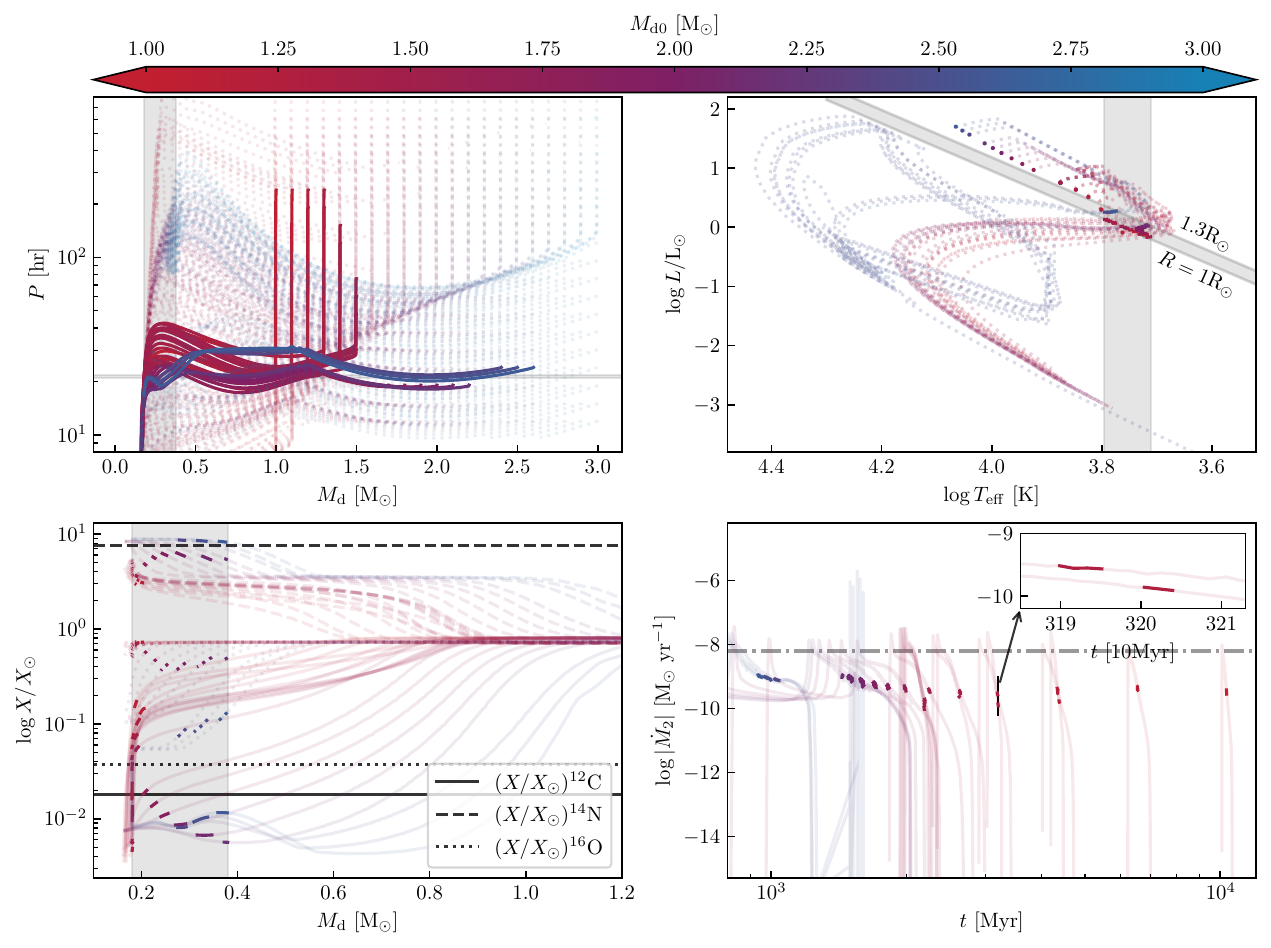}
	\caption{The binary evolutionary trajectories using the "rm12" MB law and $Z=0.014$, with colors representing the initial donor masses $M_{\rm d0}$. Those passing through the observed ranges (the donor mass $M_{\rm d}$, the orbital period $P$, the effective temperature $T_{\rm eff}$ and radius $R_{\rm d}$ of the donor) of J1858 are denoted by dark colors. The first panel (upper left) shows the full trajectories of the selected systems, while the others only depict those falling within the specified ranges. The upper two panels are similar with those in Figure~\ref{fig:example}, and the lower panels exhibit the evolution of the surface element abundances ($^{12}\rm C$, $^{14}\rm N$, $^{16}\rm O$) with $M_{\rm d}$ and the MT rates $\dot{M}_2$ with time. The horizontal lines represent the fitting values (lower left) and the upper limit (lower right), respectively.}
	\label{fig:rm12_z014}
\end{figure*}

We have evolved a number of incipient NS LMXBs with different MB prescriptions using the binary evolution code MESA (version number r22.11.1) \citep{2011ApJS..192....3P,2013ApJS..208....4P,2015ApJS..220...15P,2018ApJS..234...34P,2019ApJS..243...10P,2023ApJS..265...15J}. The NS in the LMXBs is taken as a mass point with an initial mass of 1.4 $\rm M_{\odot}$. We set the main-sequence donor to have an initial mass $M_{\rm d0}\in [1,3]\,\rm M_{\odot}$ and orbital period $\log P_0\,({\rm d})\in[-0.5,1.5]$. The size of each [$M_{\rm d0}$, $\log P_0$] grid is $0.1\times0.1$. We adopt the mixing length parameter $\alpha=2$, and the \citet{1988A&A...202...93R} scheme to compute the MT rates $\dot{M}_2$ via Roche-lobe overflow. The mass transfer efficiency onto the NS is set to 0.3 \citep{1999A&A...350..928T,2021MNRAS.503.3540C}, and the accretion rate of the NS is also limited by the Eddington accretion rate. The metallicities $Z$ for the donor stars varies from 0.01 to 0.02 in steps of 0.002. All these models are evolved until the Hubble time $\tau_{\rm Hubble}$, unless numerical divergences arise. The MB prescriptions used in our simulations are briefly introduced in the following subsections.

\subsection{The "rap83" model}\label{s:rap83}
The model proposed by \citet{1983ApJ...275..713R} (hereafter denote as "rap83") is one of the most popular MB models used in binary evolution calculations, and is also the default MB prescription in MESA. Based on the  \citet{1972ApJ...171..565S} MB law for low-mass main sequence stars, \citet{1983ApJ...275..713R} derived the MB formulation for LMXBs as follows:
\begin{equation}
\dot{J}_{\rm mb,sk}=-3.8\times10^{-30}M_{\rm d}R_{\odot}^4\left(\frac{R_{\rm d}}{R_{\odot}}\right)^{\gamma_{\rm mb}}\Omega^3\ \rm dyn\ cm, \label{eq:"rap83"}
\end{equation}
where $R_{\odot}$ is the solar radius and $\Omega$ is the angular velocity of the donor, which is assumed synchronized with the orbital angular velocity. Here we take the power law index $\gamma_{\rm mb}$ to be 4 , consistent with in the previous works \citep{2019MNRAS.483.5595V,2021ApJ...909..174D,2023ApJ...950...27G}.

\subsection{The "rm12" model}\label{s:rm12}
To better match the rotational evolution of very low-mass star ($\leq 0.5\ \rm M_{\odot}$), \citet{2012ApJ...746...43R} developed a new formulation of angular momentum evolution under the assumption of radial magnetic fields (hereafter referred to "rm12"), that is
\begin{equation}
\dot{J}_{\rm mb,rm}=-C\left[\Omega\left(\frac{R_{\rm d}^{16}}{M_{\rm d}^2}\right)^{1/3}\right]\   \rm for\ \Omega\ge \Omega_{\rm crit}, \label{eq:rm12_1}
\end{equation}
\begin{equation}
\dot{J}_{\rm mb,rm}=-C\left[\left(\frac{\Omega}{\Omega_{\rm crit}}\right)^{4}\Omega\left(\frac{R_{\rm d}^{16}}{M_{\rm d}^2}\right)^{1/3}\right]\   \rm for\ \Omega< \Omega_{\rm crit}, \label{eq:rm12_2}
\end{equation}
with
\begin{equation}
C\equiv \frac{2}{3}\left(\frac{B_{\rm crit}^{8}}{G^{2}K_{V}^{4}\dot{M}_{\rm d,w}}\right)^{1/3}. \label{eq:rm12_3}
\end{equation}
Here $G$ is the gravitational constant, $\dot{M}_{\rm d,w}$ the wind mass loss rate, $K_V$ the velocity scaling factor,  and $B_{\rm crit}$ the critical strength of magnetic field. A critical angular velocity $\Omega_{\rm crit}$ is introduced, beyond which the magnetic field reaches saturation and remains constant at $B_{\rm crit}$. In Equation (4) $\Omega_{\rm crit}$ and $C$ are taken to be $3\Omega_{\odot}=8.58\times 10^{-6}\ \rm s^{-1}$ and $2.66\times 10^3\ (\rm g^5cm^{-10}s^3)^{1/3}$, respectively.

\subsection{The "matt12" model}\label{s:matt12}
Building upon the two-dimensional axis-symmetric magnetohydrodynamic simulations conducted by \citet{2008ApJ...678.1109M}, \citet{2012ApJ...754L..26M} further performed a parameter study covering a wide range of relative magnetic field strengths and rotation rates. Subsequently, they derived a semi-analytic formula for angular momentum loss rate of Sun-like stars (hereafter "matt12"):
\begin{equation}
\dot{J}_{\rm mb,matt}=\frac{K_1^2}{(2G)^m}\bar{B}_{*}^{4m}\dot{M}_{\rm d,w}^{1-2m}\frac{R_{\rm d}^{5m+2}}{M_{\rm d}^m}\frac{\Omega}{(K_2^2+0.5u^2)^m}, \label{eq:matt12_1}
\end{equation}
where $u\equiv \Omega R_{\rm d}^{3/2}(GM_{\rm d})^{-1/2}$ is the equatorial rotation speed divided by the break-up speed of the star; $K_1=6.7$, $K_2=0.506$, and $m=0.17$ are taken from \citet{2013A&A...556A..36G}, and the stellar wind mass loss rate $\dot{M}_{\rm d,w}$ follows the prescription from \citet{1975MSRSL...8..369R},
\begin{equation}
\dot{M}_{\rm d,w}=-4\times 10^{-13}\left(\frac{R_{\rm d}}{R_{\odot}}\right)\left(\frac{L_{\rm d}}{L_{\odot}}\right)\left(\frac{M_{\odot}}{M_{\rm d}}\right)\ \rm {M_{\odot}yr^{-1}}. \label{eq:matt12_2}
\end{equation}
Here $\bar{B}_{*}=f_{*}B_{*}$ is the mean magnetic field, in which $f_{*}$ is the filling factor reflecting the magnetized
fraction of the stellar surface \citep{1996IAUS..176..237S} and $B_{*}$ is  the strength of solar surface magnetic field $B_{\odot}=1\rm G$ \citep{2021ApJ...909..174D}. \citet{2016A&A...587A.105A} recalibrated the expression of $f_*$ to match the solar mass-loss value at the age of the Sun for solid-body rotating models as:
\begin{equation}
f_*=\frac{0.4}{[1+(x/0.16)^{2.3}]^{1.22}}. \label{eq:matt12_3}
\end{equation}
Here $x=\left(\frac{\Omega_{\odot}}{\Omega}\right)\left(\frac{\tau_{\odot,\rm conv}}{\tau_{\rm conv}}\right)$ is the normalized Rossby number, and $\Omega_{\odot}\approx 3\times 10^{-6}\ \rm s^{-1}$ and $\tau_{\odot,\rm conv}\approx 2.8\times 10^6\ \rm s$ are the rotation rate and the turnover time of the convective eddies of the Sun, respectively.

\subsection{The "cboost"/"inter"/"wboost" model}\label{s:boost}
It has been noted that the "rap83" model cannot adequately reproduce the MT rates of the observed persistent NS LMXBs \citep{2002ApJ...565.1107P,2019MNRAS.483.5595V}. Therefore, \citet{2019MNRAS.483.5595V} modified the Skumanich MB law by incorporating a scaling of the magnetic field strength with the convective turnover time and a scaling of MB with the wind mass-loss rate, that is
\begin{equation}
\dot{J}_{\rm mb,boost}=\dot{J}_{\rm mb,sk}\left(\frac{\Omega}{\Omega_{\odot}}\right)^{\beta}\left(\frac{\tau_{\rm conv}}{\tau_{\odot,\rm conv}}\right)^{\xi}\left(\frac{\dot{M}_{\rm d,w}}{\dot{M}_{\odot,\rm w}}\right)^{\alpha}. \label{eq:boost}
\end{equation}
Here, the power indices $(\beta,\xi,\alpha)$ are adjustable parameters correspond to different modified MB laws, including the "Convection-boosted" (hereafter "cboost", with $\beta=0$, $\xi=2$, $\alpha=0$), the "intermediate" (hereafter "inter", with $\beta=0$, $\xi=2$, $\alpha=1$), and the "Wind-boosted" (hereafter "wboost", with $\beta=2$, $\xi=4$, $\alpha=1$) laws.

\subsection{The "carb" model}\label{s:carb}
By incorporating the relationship between the magnetic field strength and the outer convective zone, as well as the dependency of the Alfv\'en radius on the donor star's rotation, \citet{2019ApJ...886L..31V} developed the Convection And Rotation Boosted ("carb") MB model, as formulated below:
\begin{align}
\nonumber\dot{J}_{\rm mb,carb}=&-\frac{2}{3}\dot{M}_{\rm d,w}^{1/3}R_{\rm d}^{14/3}(v_{\rm esc}^2+2\Omega^2R_{\rm d}^2/K_{2,*}^{2})^{-2/3}\\
& \times\Omega_{\odot}B_{\odot}^{8/3}\left(\frac{\Omega}{\Omega_{\odot}}\right)^{11/3}\left(\frac{\tau_{\rm conv}}{\tau_{\odot,\rm conv}}\right)^{8/3}, \label{eq:"carb"}
\end{align}
where $v_{\rm esc}$ is the surface escape velocity and $K_{2,*}=0.07$ is obtained from the simulations conducted by \citet{2015ApJ...798..116R}, which sets the limit where the rotation rate begins to play a significant role.

Finally we notice that the conditions when MB works in MESA version r22.11.1 differ from those in the older versions used in previous studies \citep{2019MNRAS.483.5595V,2019ApJ...886L..31V,2021ApJ...909..174D,2023ApJ...950...27G}, which only check if a radiative core exist. In MESA r22.11.1, the MB criteria are more strict, described as follows: the mass fraction of the convective envelope should be larger than $10^{-6}$ and less than 0.99, and the mass fraction of the convective core has to be less than 0.02.

\section{The evolution of NS LMXBs}\label{s:result}
\begin{figure*}[t]
	\centering \includegraphics[width=\textwidth]{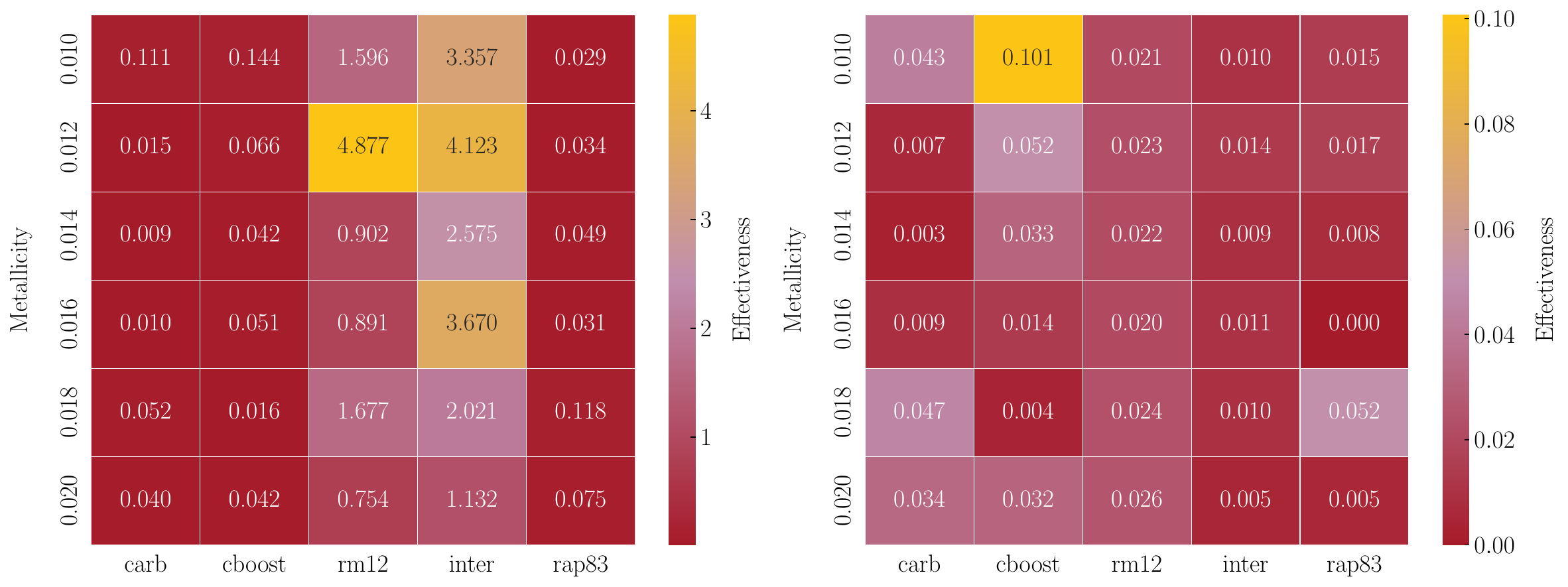}
	\caption{The model effectiveness for different MB law and metallicity combinations. The left panel takes account of the observed ranges of $M_{\rm d}$ and $P$ only, while the right one additionally incorporates the constraints on $T_{\rm eff}$ and $R_{\rm d}$.}
	\label{fig:a1858}
\end{figure*}

\begin{figure*}[t]
	\centering \includegraphics[width=\textwidth]{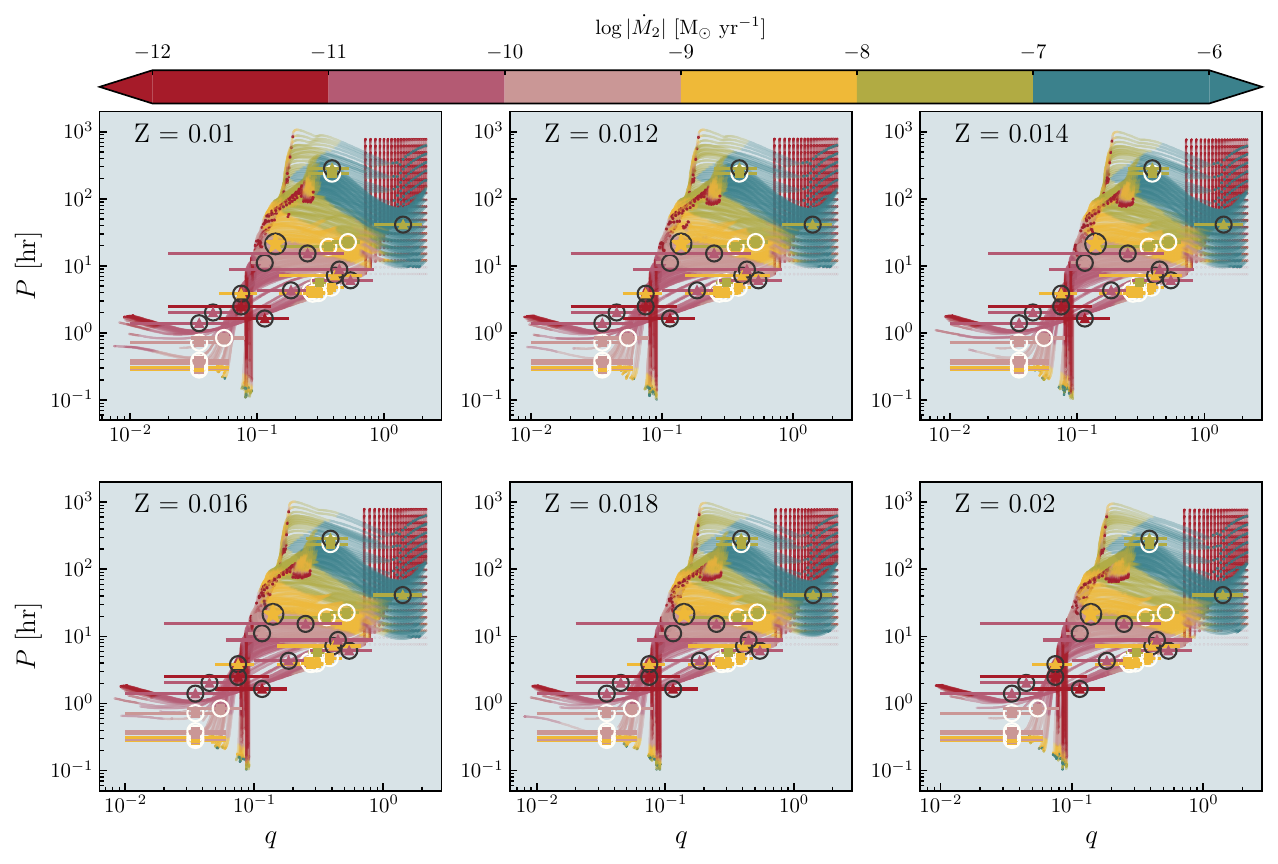}
	\caption{The evolution of LMXBs using "rm12" MB law with initially different mass ratio $q=M_{\rm d}/M_{\rm NS}$, $P$ and $Z$. Different colors signify the magnitudes of MT rates. The squares with the white circles represent persistent LMXBs and UCXBs, and the triangles with the black circles denote transient LMXBs.The circled star represents J1858.}
	\label{fig:rm12}
\end{figure*}

\begin{figure}[t]
	\centering \includegraphics[width=.45\textwidth]{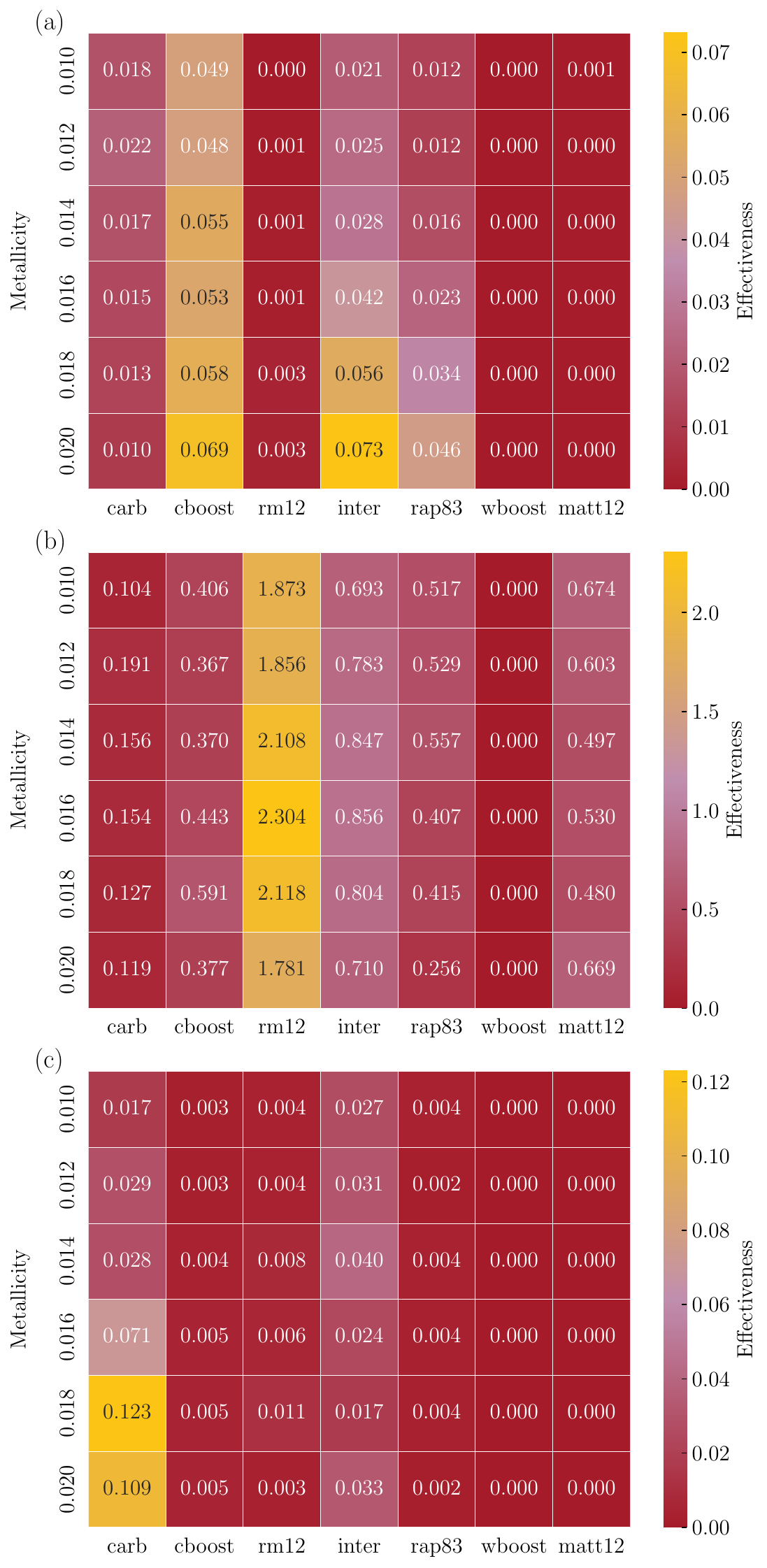}
	\caption{The assessed effectiveness for different models of MB and metallicity. Panels a, b and c represent the results for persistent, transient LMXBs, and UCXBs, respectively.}
	\label{fig:norm}
\end{figure}
\subsection{The case of J1858}\label{s:j1858}

J1858 is a transient NS LMXB firstly detected with \textit{Neil Gehrels Swift Observatory} \citep{2004ApJ...611.1005G} on October 2018 \citep{2018ATel12151....1K}. As stated in Section~\ref{s:intro}, there exists discrepancy in the estimated donor mass $M_{\rm d}$: observations by \cite{2022MNRAS.514.1908K} indicate $M_{\rm d}\approx 0.2\ \rm M_{\odot}$, but simulations by \cite{2024MNRAS.527.2508C} with the "carb" MB recipe and $Z=0.02$ metallicity suggest $0.5\ \rm{M}_{\odot}\leq$ $M_{\rm d}\leq 1.3\ \rm{M}_{\odot}$.

We firstly investigate the influence of the MB laws and metallicity on the evolution of this source with initial parameters $M_{\rm d0}=1.5\,\rm M_{\odot}$ and $P_0=1.5\,\rm d$. The left and right panels of Figure~\ref{fig:example} depict the evolution in the $M_{\rm d}-P$ plane and the Hertzsprung-Russell (HR) diagram, respectively. And the shaded areas represent the observed (or inferred) values of J1858. We constrain the current mass and radius of the donor within the range of $0.18\ \rm{M}_{\odot}\leq$ $M_{\rm d}\leq 0.38\ \rm{M}_{\odot}$ and $1\ \rm{R}_{\odot}\leq$ $R_{\rm d}\leq 1.3\ \rm{R}_{\odot}$, respectively \citep{2022MNRAS.514.1908K}. The present orbital period and effective temperature are $P=21.3\pm 0.5\ \rm hrs$ and $5140\ \rm K\leq$ $T_{\rm eff}\leq 6240\ \rm K$ \citep{2024MNRAS.527.2508C}. It is apparent that different MB prescriptions and $Z$ result in markedly diverse trajectories in both panels, even when the initial $M_{\rm d}$ and $P$ are identical. In general, the majority of the MB laws tend to produce converging systems, three of which are expected to drive the binary pass through the observed ranges under these initial conditions. The efficiencies of the "rap83" and "matt12" MB laws appear to be weaker than others, while those of "inter" and "wboost" recipes are relatively stronger. Moreover, a lower $Z$ appears to yield a wider orbit during the evolution under the same MB recipe, except for the "matt12" law. As a result, taking the "carb" MB law as an example, the evolutionary trajectory with $Z=0.01$ can simultaneously pass through the observed ranges of $M_{\rm d}$ and $P$, whereas it fails with the other two metallicities.

We then evolve with multiple grids of the input parameters to search for the possible initial conditions which can satisfy the observational properties of J1858 (i.e., $M_{\rm d}$, $P$, $T_{\rm eff}$ and $R_{\rm d}$). In Figure~\ref{fig:rm12_z014}, we illustrate the simulated results using the "rm12" MB law and $Z=0.014$, where the colors denote the initial donor masses $M_{\rm d0}$. The evolutionary trajectories passing through the observed ranges of J1858 are dyed with dark colors in the figure (The evolutionary trajectories with other MB laws are presented in Figures~\ref{fig:inter_z014}-\ref{fig:rap83_z014} in the Appendix). Overall, there are totally 22 selected trajectories passing through the observational ranges of J1858.  The first (upper left) panel illustrates the full trajectories of the selected systems, while the others only depict those falling within the specified ranges. From the first panel, we observe that adopting the initial parameters with $M_{\rm d0}\lesssim 1.5\,\rm M_{\odot}$ and $P_0\sim 1-10$ d, as well as with $M_{\rm d0}> 1.5\,\rm M_{\odot}$ and $P_0\sim 1$ d can meet the observational constraints. And the systems with low $M_{\rm d0}$ may eventually evolve into ultra-compact X-ray binaries (UCXBs), typically characterized by orbital periods shorter than 80 minutes and donor stars that are partially or completely degenerate. According to the related position in the HR diagram,  it can be inferred that J1858 is likely undergoing the early case-B MT\footnote{Case A, B and C mass transfer refer to the Roche lobe overflow process when the donor is undergoing core hydrogen burning, shell hydrogen burning and stages after exhaustion of the core helium burning, respectively.}. Moreover, \citet{2024MNRAS.527.2508C} modeled the ultraviolet line ratios of J1858 and estimated the equilibrium surface abundances of the donor to be $(X/X_{\odot})_{^{12}\rm C}\approx 0.018$, $(X/X_{\odot})_{^{14}\rm N}\approx 7.5$ and $(X/X_{\odot})_{^{16}\rm O}\approx 0.037$, as shown with the horizontal lines in the lower left panel of Figure~\ref{fig:rm12_z014}. We notice that the simulated surface abundances of $^{12}$C and $^{14}$N closely match the measured values, and the $^{16}$O abundance is slightly higher, particularly in the case of lower $M_{\rm d0}$. Additionally, we present the evolution of $\dot{M}_2$ with the donor age in the lower right panel. The duration of the evolutionary stages which match the observations can persist for several to tens of million years, with $\dot{M}_2$ well below
$10^{-8.2}\ \rm M_{\odot}\ yr^{-1}$, as constrained by its transient behavior.

We further explore whether alternative MB laws and metallicities can effectively account for the observations of J1858. We define the parameter $A_i=\tau_{i}/\tau_{\rm Hubble}$ to assess the effectiveness of a specific (MB$+Z$) model, where $\tau_{i}$ denotes the duration for which the evolution of the $i$-th incipient binary meets the observational constraints of J1858 and $\tau_{\rm Hubble}=14\,\rm Gyr$ is Hubble time. Then, for the 441 grids of systems that we have evolved for each model, we calculate 
\begin{equation}
 A_{\rm J1858}=\sum\limits_{i=1}^{441} A_{i}  
\end{equation} 
to evaluate the metric of each model, with a higher value indicating greater effectiveness for J1858. In Figure~\ref{fig:a1858}, we illustrate the calculated $A_{\rm J1858}$ for various models, where the left panel exclusively considers the allowed ranges of $M_{\rm d}$ and $P$, and the right one additionally incorporates the constraints on $T_{\rm eff}$ and $R_{\rm d}$. None of the trajectories using the "matt12" and "wboost" laws can match the observations, and thus are not shown. In the left panel, the "rm12" and "inter" laws with $Z=0.012$ seem to be most promising to account for the ranges of $M_{\rm d}$ and $P$ of J1858; in the right panel, the "cboost" law with $Z=0.01$ shows the superior performance when taking $T_{\rm eff}$ and $R_{\rm d}$ into account, mainly because of more extended duration of the evolutions that match the observational constraints.  Moreover, lower $Z$ can considerably decrease the equilibrium abundances, especially that of $^{16}\rm O$.  Generally, systems with initial $Z\leq 0.014$ and $M_{\rm d0}\gtrsim 2\ \rm M_{\odot}$ seem to better align with the derived abundances by \citet{2024MNRAS.527.2508C}.

\subsection{Comparison with other LMXBs}\label{s:lmxb}

\begin{table*}[]
\setlength{\tabcolsep}{3mm}
    \caption{The effectiveness of different MB laws with initial $Z=0.014$ for the observed LMXBs. From top to bottom, the three categories of sources are persistent and transient LMXBs, and UCXBs, respectively.}
	\label{tab:z014}
    \centering
	\begin{tabular}{c|ccccccc}
		\hline \hline
		Source name       & carb     & cboost   & rm12     & inter    & rap83    & matt12   & wboost\\         \hline
		4U 1636-536       & 0.00E+00 & 6.53E-03 & 0.00E+00 & 1.31E-02 & 1.03E-02 & 0.00E+00 & 0.00E+00\\
		GX 9+9            & 0.00E+00 & 1.58E-02 & 0.00E+00 & 9.98E-04 & 0.00E+00 & 0.00E+00 & 0.00E+00\\
		4U 1735-444       & 0.00E+00 & 1.21E-02 & 0.00E+00 & 4.39E-03 & 0.00E+00 & 0.00E+00 & 0.00E+00\\
		2A 1822-371       & 1.56E-04 & 0.00E+00 & 0.00E+00 & 0.00E+00 & 0.00E+00 & 0.00E+00 & 0.00E+00\\
		Sco X-1           & 8.94E-03 & 5.34E-03 & 0.00E+00 & 1.52E-03 & 0.00E+00 & 0.00E+00 & 0.00E+00\\
		GX 349+2          & 7.23E-03 & 1.18E-02 & 9.97E-04 & 0.00E+00 & 0.00E+00 & 0.00E+00 & 0.00E+00\\
		Cyg X-2           & 1.11E-03 & 1.25E-03 & 2.75E-05 & 0.00E+00 & 2.95E-04 & 3.30E-04 & 0.00E+00\\
		4U 1254-69        & 0.00E+00 & 2.36E-03 & 0.00E+00 & 8.16E-03 & 5.59E-03 & 0.00E+00 & 0.00E+00\\ \hline
		HETE J1900.1-2455 & 0.00E+00 & 9.15E-02 & 4.66E-02 & 1.16E-01 & 4.98E-02 & 0.00E+00 & 0.00E+00\\
		1A 1744-361       & 2.47E-02 & 3.01E-02 & 7.44E-02 & 5.41E-02 & 2.10E-02 & 0.00E+00 & 0.00E+00\\
		SAX J1808-3658    & 5.20E-02 & 1.27E-01 & 2.56E-01 & 5.06E-01 & 4.92E-02 & 0.00E+00 & 0.00E+00\\
		IGR 00291+5394    & 1.34E-02 & 1.70E-02 & 3.83E-02 & 6.14E-02 & 0.00E+00 & 0.00E+00 & 0.00E+00\\
		EXO 0748-676      & 8.56E-03 & 0.00E+00 & 0.00E+00 & 0.00E+00 & 0.00E+00 & 0.00E+00 & 0.00E+00\\
		XTE J1814-338     & 0.00E+00 & 0.00E+00 & 4.26E-01 & 0.00E+00 & 0.00E+00 & 4.47E-03 & 0.00E+00\\
		XTE J2123-58      & 1.42E-04 & 0.00E+00 & 0.00E+00 & 0.00E+00 & 0.00E+00 & 1.71E-01 & 0.00E+00\\
		X 1658-298        & 0.00E+00 & 2.58E-02 & 0.00E+00 & 3.06E-02 & 1.44E-02 & 0.00E+00 & 0.00E+00\\
		SAX J1748.9-2021  & 1.30E-02 & 9.11E-03 & 9.02E-01 & 5.02E-02 & 2.98E-01 & 3.11E-01 & 0.00E+00\\
		IGR J18245-2452   & 1.75E-02 & 2.34E-02 & 2.91E-01 & 1.93E-02 & 0.00E+00 & 0.00E+00 & 0.00E+00\\
		Cen X-4           & 2.48E-02 & 4.39E-02 & 7.39E-02 & 9.90E-03 & 1.20E-01 & 0.00E+00 & 0.00E+00\\
		Her X-1           & 4.69E-04 & 5.58E-04 & 2.15E-05 & 1.64E-05 & 1.48E-03 & 6.10E-03 & 0.00E+00\\
		GRO J1744-28      & 1.60E-03 & 9.46E-04 & 2.00E-05 & 0.00E+00 & 2.69E-03 & 4.63E-03 & 0.00E+00\\ \hline
		4U 0513-40        & 8.74E-04 & 2.89E-04 & 6.05E-04 & 3.52E-03 & 3.12E-04 & 0.00E+00 & 0.00E+00\\
		2S 0918-549       & 1.05E-03 & 3.65E-04 & 7.21E-04 & 4.11E-03 & 4.73E-04 & 0.00E+00 & 0.00E+00\\
		4U 1543-624       & 1.50E-03 & 5.42E-04 & 9.07E-04 & 5.46E-03 & 6.98E-04 & 0.00E+00 & 0.00E+00\\
		4U 1850-087       & 3.40E-03 & 1.60E-03 & 1.95E-03 & 1.10E-02 & 1.38E-03 & 0.00E+00 & 0.00E+00\\
		M15 X-2           & 2.99E-03 & 1.48E-03 & 3.42E-03 & 1.63E-02 & 1.06E-03 & 0.00E+00 & 0.00E+00\\
		4U 1626-67        & 9.74E-03 & 0.00E+00 & 0.00E+00 & 0.00E+00 & 0.00E+00 & 0.00E+00 & 0.00E+00\\
		4U 1916-053       & 8.07E-03 & 0.00E+00 & 0.00E+00 & 0.00E+00 & 0.00E+00 & 0.00E+00 & 0.00E+00\\ \hline \hline
	\end{tabular}
\end{table*}

In the previous subsection, we demonstrate that the "rm12" and "inter" MB models and the "cboost" and "carb" MB models are able to account for the observations of J1858 to different extent. Naturally, we would also like to investigate whether the conclusion could apply to other LMXBs. In Figure~\ref{fig:rm12}, we depict the evolution of LMXBs with different initial $q$, $P$ and $Z$, employing the "rm12" MB prescription as a reference. The colors represent the magnitudes of the MT rates. The observed quantities of persistent and transient LMXBs are collected from Tables 4 and 5 of \citet{2019MNRAS.483.5595V}\footnote{Note that the source 4U 1254$-$69 listed in Table 5 of \citet{2019MNRAS.483.5595V} should be categorized as a persistent source, not a transient one.}, denoted by squares with white circles and triangles with black circles, respectively, and the star represents J1858. There are not remarkable differences in the evolutions of LMXBs with different $Z$, suggesting that its influence is insignificant. The majority of LMXBs can be covered in the $P-q$ space. Although the MT rates of the persistent sources seem not to be fully matched, the MT rates of most transient sources and UCXBs can be well reproduced.

In Table~\ref{tab:z014}, we compare the effectiveness for the observed LMXBs under different MB prescriptions with $Z=0.014$. The persistent and transient sources are listed in the upper and middle parts, respectively. The sources in the lower part are UCXBs, which can be effectively reproduced using the "carb" MB law. Other MB laws, apart from "matt12" and "wboost", are also able to reproduce the observed UCXBs except 4U 1626$-$67 and 4U 1916$-$053. For other persistent sources, the "cboost" MB law scores the highest, while the "inter" and "carb" MB laws also perform well. On the other hand, the "inter" and "rm12" laws exhibit excellent performances in reproducing the observed transient sources overall. The simulations employing the "carb" and "cboost" laws can also account for the majority of the systems, even for those which cannot be explained by the "rm12" and "inter" laws. Moreover, the "matt12" MB law seems to be capable of explaining the transient sources with relatively long orbital periods, such as Her X$-$1 and GRO J1744$-$28.

We then assess the effectiveness $A$ of each model with different combinations of MB prescription and metallicity for the known LMXB population, similar as we have done for J1858. We classify LMXBs into persistent (without UCXBs), transient LMXBs, and UCXBs, and present the results for each class in Figure~\ref{fig:norm}a, b, and c, respectively. Overall, the "cboost" and "inter" MB laws with higher $Z$ are more promising for the persistent sources, while the "rm12" and "inter" laws outperform others for the transient sources. Additionally, the "carb" MB law with higher $Z$ score the highest for UCXBs.

\section{Discussion and Conclusions}\label{s:conclusion}
\subsection{Influence on the bifurcation periods}\label{s:bifurcation}
\begin{figure}[t]
	\centering \includegraphics[width=.45\textwidth]{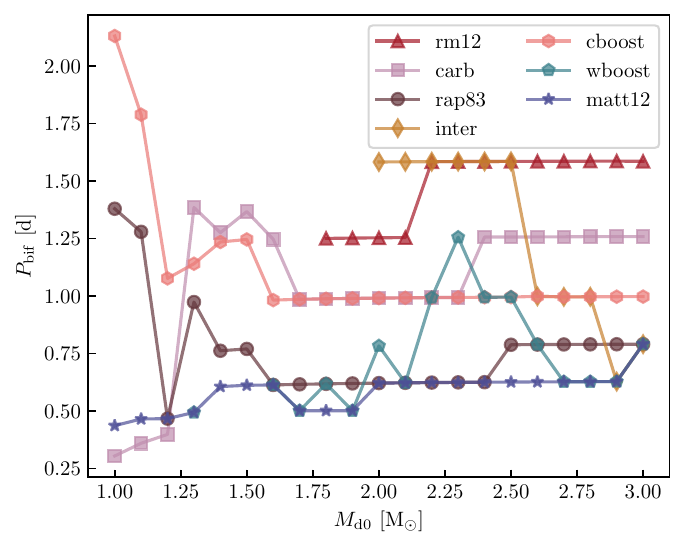}
	\caption{Comparisons of the bifurcation periods $P_{\rm bif}$ for different MB laws with $z=0.02$.}
	\label{fig:pbif_z02}
\end{figure}

The bifurcation period $P_{\rm bif}$ is a critical orbital period distinguishing the LMXBs evolving towards converging systems from those towards diverging systems \citep[e.g.,][]{1988A&A...191...57P,1989A&A...208...52P}. Its magnitude was shown to be highly dependent on the adopted MB models \citep{2009ApJ...691.1611M,2021ApJ...909..174D}. Here we define $P_{\rm bif}$ as the orbital period when Roche lobe overflow just begins \citep{2002ApJ...565.1107P}. In Figure~\ref{fig:pbif_z02}, we illustrate $P_{\rm bif}$ as a fuction of $M_{\rm d}$ with $Z=0.02$, using different MB laws. When $M_{\rm d0}\ge 1.7\ \rm M_{\odot}$, the $P_{\rm bif}$ values predominantly fall within the range of $[0.5,1.5]$ d and remain unaffected by $M_{\rm d0}$.
However, the evolution of $P_{\rm bif}$ when $M_{\rm d0}\lesssim 1.7 \ \rm M_{\odot}$ becomes more complicated. The absence of $P_{\rm bif}$ at lower $M_{\rm d}$ with the "inter", "rm12" and "wboost" MB laws means that $P_{\rm bif}$ exceeds the maximum initial period $10^{1.5}$ days because of the strong MB strengths. In comparison, the MB strengths of the "matt12" and "rap83" MB laws are relatively weaker, resulting in lower $P_{\rm bif}$ values.

\subsection{Comparisons with other works}\label{s:compare}
\begin{figure*}[t]
	\centering \includegraphics[width=\textwidth]{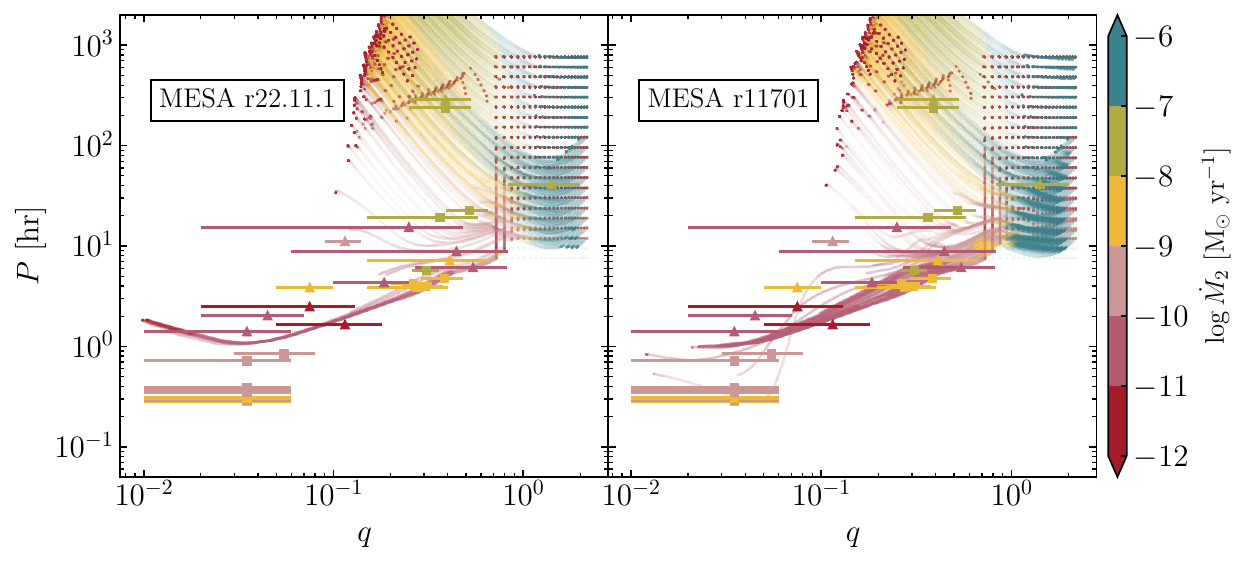}
	\caption{Comparisons of evolutionary trajectories between MESA versions r22.11.1 and r11701 with $Z=0.014$ by employing the MB law proposed by \citet{2018ApJ...862...90G}.}
	\label{fig:gar18}
\end{figure*}

There have been several related works on the effects of the MB laws on the evolution of LMXBs \citep{2019MNRAS.483.5595V,2019ApJ...886L..31V,2021ApJ...909..174D,2023ApJ...950...27G,2024MNRAS.530.4277E}. Similar to  \citet{2019ApJ...886L..31V}, \citet{2019MNRAS.483.5595V} and \citet{2021ApJ...909..174D}, we find that the "cboost", "inter" and "carb" laws better reproduce persistent LMXBs than other MB laws. In addition, we show that UCXBs can be well reproduced with the "carb" MB law, aligning with  \citet{2019ApJ...886L..31V}. However, different from \citet{2019MNRAS.483.5595V} and \citet{2021ApJ...909..174D}, we find that the "inter" and "cboost" MB laws are not preferred for UCXBs, especially for 4U 1626$-$67 and 4U 1916$-$053. Moreover, for transient LMXBs, we find that the "rm12", "inter", and "cboost" MB laws all preform well, which are not explored in previous works.

The discrepancies may partly originate from different conditions for the occurrence of MB in different versions of MESA. We have used the newer version r22.11.1 in this paper while most previous works employed version 11554 or the former. In the newer version the conditions when MB works has been considerably modified as described in Section~\ref{s:method}. For example, \citet{2023ApJ...950...27G} demonstrated that the MB law proposed by \citet{2018ApJ...862...90G} performed well in modeling stellar spin evolution in both the single- and binary-regimes by using MESA r11701. However, this MB law appears not to be accountable for the formation of most LMXBs when using MESA r22.11.1, particularly for UCXBs, as depicted in Figure~\ref{fig:gar18}. \citet{2024MNRAS.530.4277E}  investigated the effect of the MB laws on the evolution of LMXBs with their own stellar code \citep{2003MNRAS.342...50B}, in which they assumed that MB fully acts if the ratio of the outer convective zone mass and the stellar mass is larger than $10^{-2}$.

Moreover, the previous works have predominantly focused on the metallicity of either $Z=0.02$ or 0.0142 \citep{2009ARA&A..47..481A}. In our paper, we consider diffenent metallicities to investigate their impact on the evolution of LMXBs.

\subsection{Summary}\label{s:summary}
To resolve the controversy in the donor mass of the transient NS LMXB J1858 derived with different methods, we conduct simulations of LMXB evolution with various kinds of MB laws and metallicities. We find that employing the "rm12" and "inter" MB laws, and the "cboost" or "carb" MB laws may effectively account for the observations of J1858 to different extent, with the lifetime satisfying the observational constraints persisting for several million years or longer.

Furthermore, we investigate whether the conclusion applies to other observed LMXBs. We find that among the seven selected MB laws, the "rm12" and "inter" MB laws are the most promising for transient LMXBs. Meanwhile, the "cboost" MB prescription also exhibits satisfactory performance. However, the simulations with "rm12" MB law fails to reproduce the high MT rates in nearly half of the persistent LMXBs, while the "cboost" and "inter" laws might be more favorable for persistent LMXBs, similar to the conclusions drawn by \citet{2019ApJ...886L..31V} and \citet{2021ApJ...909..174D}. Regarding UCXBs, the "carb" MB law appears to be the most promising explanation.

It should be mentioned that when comparing with the observations, we have only considered uniformly distributed $M_{\rm d0}$ and $P_0$ and the resulting trajectories passing through the parameter space of the observed LMXBs, without taking into account the formation rates of the selected incipient LMXBs. In addition, our comparison with observations is subject to observational bias and selection effects. Nevertheless, the conclusions drawn above still can offer a useful perspective for examining the impact of the MB laws and metallicities on the evolution of LMXBs, suggesting that it is essential to establish a comprehensive and unified MB model. Alternatively, the difference of MB laws in persistent and transient NS LMXBs could partly originate from different extent of X-ray irradiation on the donor star in the two types of systems. This will be the subject of our future investigation. 

We also explore the influence of different MB laws on the bifurcation periods. The results show that the "rm12" and "inter" laws typically exhibit longer $P_{\rm bif}$ attributed to their stronger MB strength. The corresponding $P_{\rm bif}$ values primarily fall within the range of $[0.5,1.5]$ d and remain unaffected by $M_{\rm d0}$ when $M_{\rm d0}\ge 1.7\ \rm M_{\odot}$, while those with lighter donor exhibit more complicated behavior.

\section*{acknowledgments}

We are grateful to the referee for helpful comments. This work was supported by the National Key Research and Development
Program of China (2021YFA0718500), the Natural Science
Foundation of China under grant Nos. 12041301 and 12121003.

\section*{Data Availability}
The MESA code, the inlist files necessary to reproduce our simulations are available at
\dataset[zenodo.13626633]{https://doi.org/10.5281/zenodo.13626633}. The other data and codes underlying this article will be shared on reasonable request to the authors.

\bibliography{YL24}{}
\bibliographystyle{aasjournal}

\appendix
\setcounter{figure}{0}
\renewcommand{\thefigure}{A\arabic{figure}}

\section{More results for J1858}
\begin{figure*}[ht]
	\centering \includegraphics[width=\textwidth]{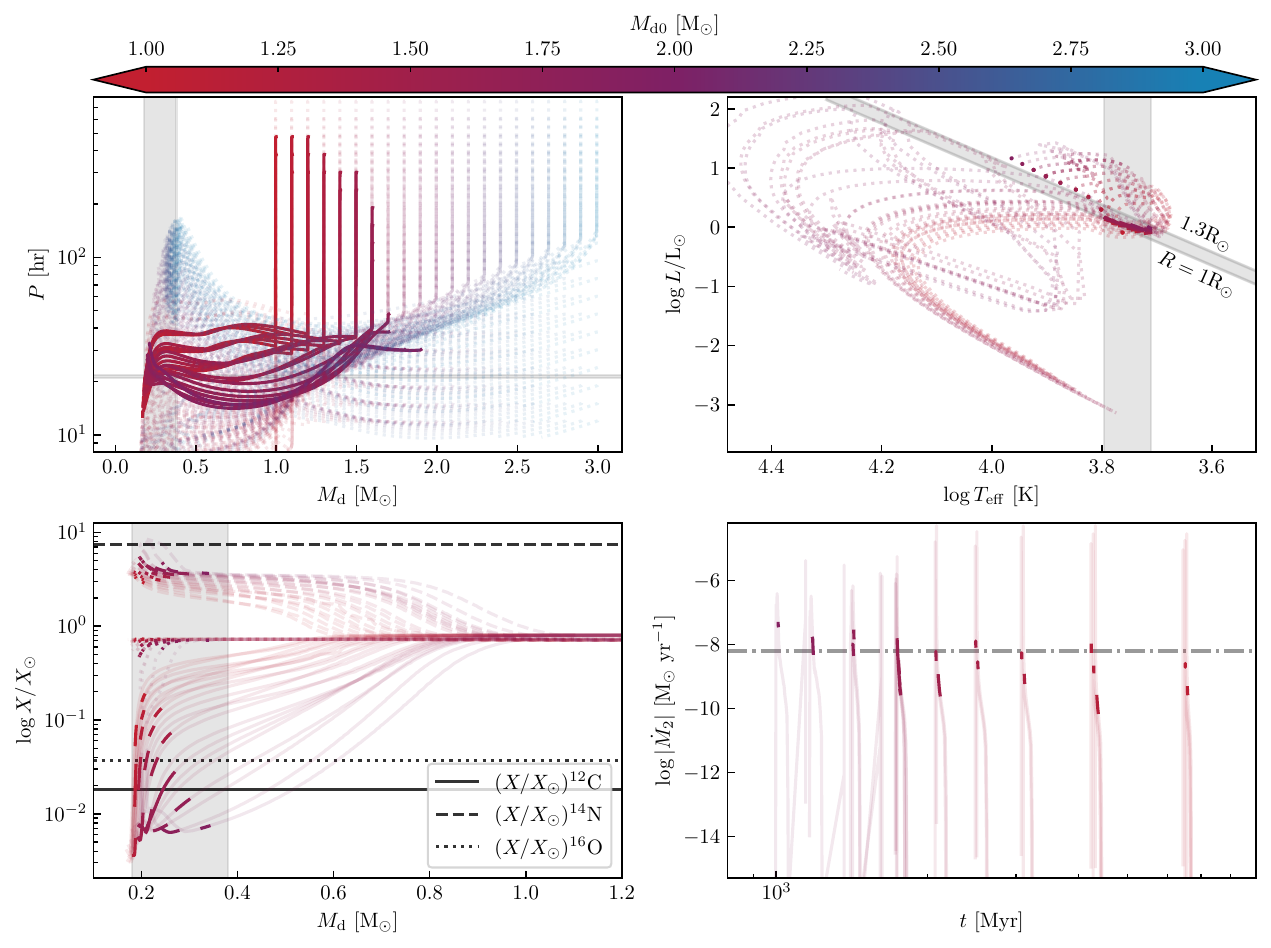}
	\caption{Similar with Figure~\ref{fig:rm12_z014} in the main text, but using "inter" MB law with $Z=0.014$.}
	\label{fig:inter_z014}
\end{figure*}

\begin{figure*}[t]
	\centering \includegraphics[width=\textwidth]{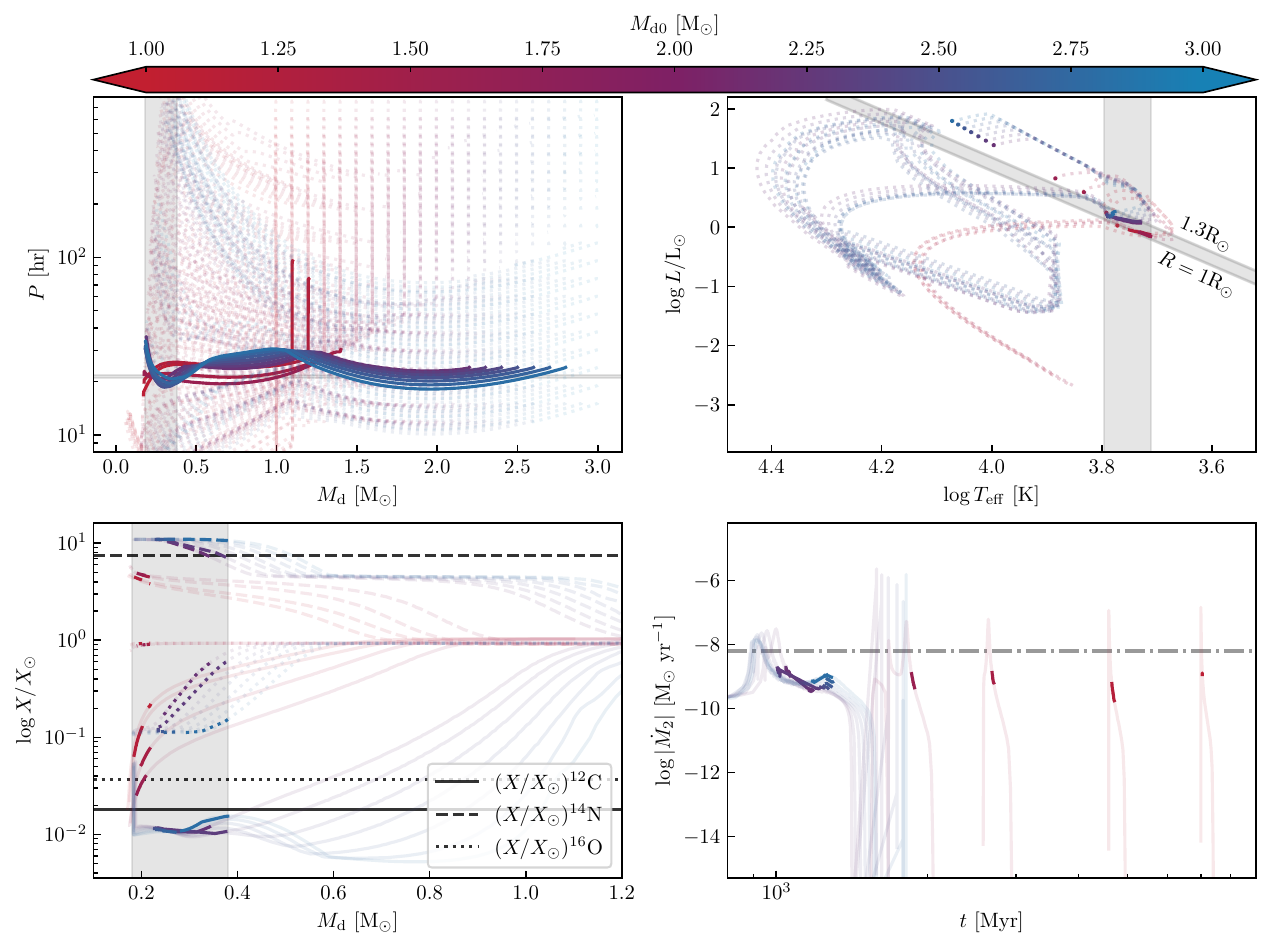}
	\caption{Similar with Figure~\ref{fig:rm12_z014} in the main text, but using "carb" MB law with $Z=0.018$.}
	\label{fig:carb_z018}
\end{figure*}

\begin{figure*}[t]
	\centering \includegraphics[width=\textwidth]{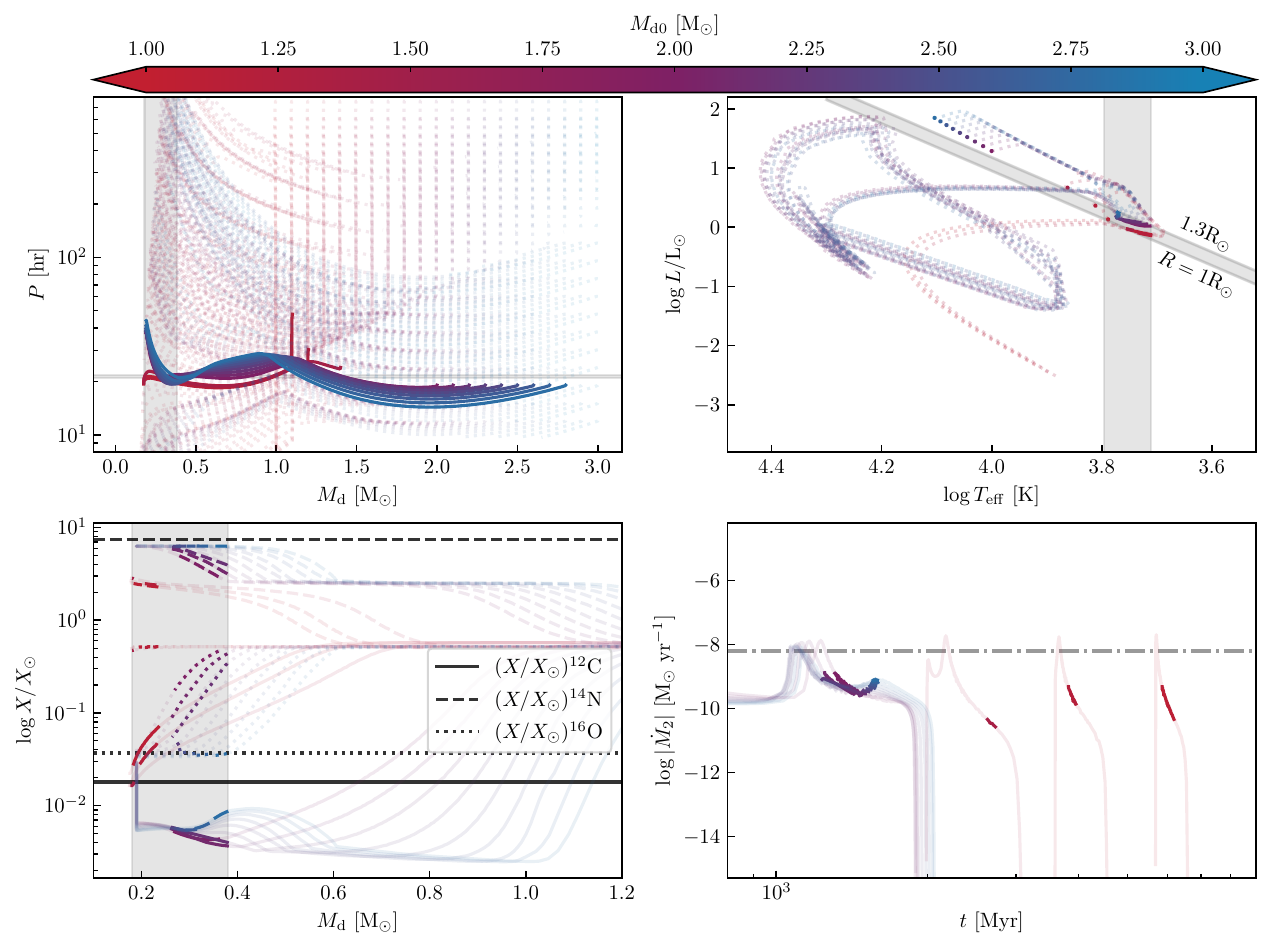}
	\caption{Similar with Figure~\ref{fig:rm12_z014} in the main text, but using "cboost" MB law with $Z=0.01$.}
	\label{fig:cboost_z01}
\end{figure*}

\begin{figure*}[t]
	\centering \includegraphics[width=\textwidth]{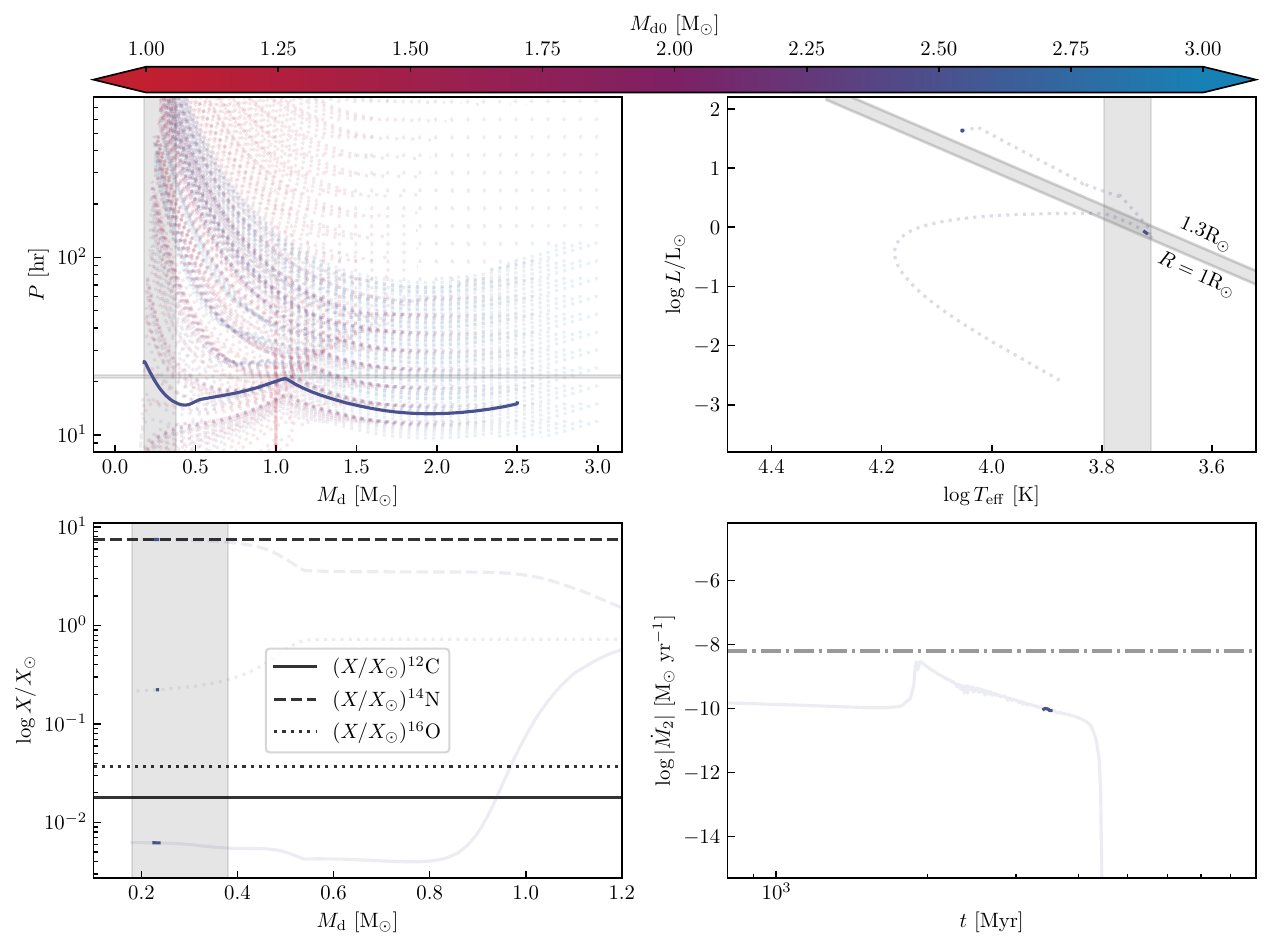}
	\caption{Similar with Figure~\ref{fig:rm12_z014} in the main text, but using "rap83" MB law with $Z=0.014$.}
	\label{fig:rap83_z014}
\end{figure*}

In the appendix, we present the results similar with Figure~\ref{fig:rm12_z014}, but using other prescriptions for MB laws and metallicities. The evolutionary trajectories simulated by using "inter" MB law with $Z=0.014$, "cboost" MB law with $Z=0.01$, "carb" MB law with $Z=0.018$ and "rap83" MB law with $Z=0.014$ are illustrated in Figures~\ref{fig:inter_z014}-\ref{fig:rap83_z014}, respectively. It is apparent that more trajectories pass through the observational intervals of J1858 with "inter" MB law, while the durations with "carb" or "cboost" MB laws are much longer. The simulation under the "rap83" MB law reveals only one qualified trajectory, as depicted in Figure~\ref{fig:rap83_z014}. No suitable systems for J1858 are simulated with the "wboost" and "matt12" MB laws, and thus are not shown.

\end{document}